\documentclass{llncs}
\usepackage{hyperref,xcolor,color,colortbl,graphicx,tabularx}
\usepackage{amssymb , pifont}
\newcommand{\xmark}{\ding{55}}\usepackage{etoolbox}
\usepackage[para]{threeparttable}
\usepackage{pgfplots}
\usepackage{multirow}
\usepackage{xspace}
\usepackage{cite}
\usepackage[acronym]{glossaries}
\usepackage{booktabs}

\usepackage{tikz}
\usetikzlibrary{shapes,arrows,shadows}
\usepackage{amsmath,bm,times}
\usepackage{environ}

\makeatletter
\newsavebox{\measure@tikzpicture}
\NewEnviron{scaletikzpicturetowidth}[1]{  \def\tikz@width{#1}  \def\tikzscale{1}\begin{lrbox}{\measure@tikzpicture}  \BODY
  \end{lrbox}  \pgfmathparse{#1/\wd\measure@tikzpicture}  \edef\tikzscale{\pgfmathresult}  \BODY
}

\newcommand{\matador}{MATAdOR\xspace}

\newcommand{\Mmss}{Mobile messaging services\xspace}
\newcommand{\mmss}{mobile messaging services\xspace}

\newcommand{\mms}{mobile messaging service\xspace}

\newcommand{\mmas}{mobile messaging applications\xspace}

\newcommand{\mma}{mobile messaging application\xspace}
\newcommand{\npath}{network path\xspace}
\newcommand{\apath}{application path\xspace}
\newcommand{\NPath}{Network Path\xspace}
\newcommand{\APath}{Application Path\xspace}

\title{Analyzing Locality of Mobile Messaging Traffic using the MATAdOR Framework}

\author{Quirin Scheitle \and Matthias Wachs\and Johannes Zirngibl\and Georg Carle \\}
\institute{Technical University of Munich (TUM)\\
Department of Informatics\\
Chair for Networking Services and Architectures \\
\email{Email: \{scheitle, wachs, carle\}@net.in.tum.de, \{zirngibl\}@in.tum.de} }

\begin{document}

\maketitle

\begin{center}
\textbf{This is a preprint of the publication to appear at PAM 2016. \newline
Last update: 14 Jan 2016.}
\end{center}

\begin{abstract}
\Mmss have gained a large share in global
telecommunications. Unlike conventional services like phone
calls, text messages or email, they do not feature a standardized environment
enabling a federated and potentially local service architecture.
We present an extensive and large-scale analysis of communication patterns
for four popular \mmss between 28 countries and analyze the locality
of communication and the resulting impact on user privacy. We show that server
architectures for \mmss are highly centralized in single countries. This forces 
messages to drastically deviate from a direct communication path, enabling
hosting and transfer countries to potentially intercept and censor traffic.
To conduct this work, we developed a measurement framework to
analyze traffic of such \mmss. It allows to conduct automated
experiments with \mmas, is transparent to those applications and does
not require any modifications to the applications.

\end{abstract}

\section{Introduction}
\Mmss like WeChat or WhatsApp see a steady
increase in both active users and messages sent, with a particular success in
emerging markets like China, Brazil or Malaysia~\cite{pew2015,2015tnsim}.
Some researchers predict a shift in communication paradigms
with \mmss eradicating classical forms
of electronic communication like email or text messages.
As an example, the number of text messages sent in Germany shrunk by $62\%$ from 2012 to 2014
\cite{bnetzajahresbericht2014}, after it had been growing exponentially for over a decade.

\Mmss and their design strongly differ from classic Internet
communication services: established means of communication---like email,
internet telephony or instant messaging---often rely on federated or
decentralized architectures, with operators providing services to their
customers and from within their domain.

\Mmss tend to abandon established principles of openness and
federation: messaging services are often realized in a closed, non-federated,
cloud-centric environment built upon proprietary communication and
security protocols neither standardized nor disclosed to the public.

This paradigm shift puts at risk the user's freedom and access to secure, confidential
and privacy-preserving communication. With such services, the user---relating to 
her social network through such applications---strongly depends
on the service provider to not modify or restrict the service. The user's privacy also depends on the legislation the operating company is
subject to: governments are often interested in controlling Internet
services~\cite{moxie2013,vodafone2015} and accessing messages~\cite{apple2015}
as well as metadata. The matters of security and privacy move along the same
lines and generally involve a full trust into a closed system, a misleading
assumption as we saw with WhatApp's announced \textit{end-to-end-encryption}, which
is supported on Android, but not Apple devices~\cite{whatsapp2014}, without giving  
feedback on encryption status to the user. 
First attempts to analyze the security properties of \mmss have for
example been made by the EFF with its \textit{Secure Messaging Scorecard}~
\cite{eff2014}.

In this work, we analyze the implications of \mmss on the users and
their privacy. Similar to the discussion about a ``nation-centric
Internet''\cite{wahlisch2012exposing}, we set out to understand the
communication behavior and patterns of \mmss by analyzing how
\textit{local} messaging traffic is from a geographic and legal point of view.
We analyze how messaging traffic is routed through the Internet and which
countries could therefore access this traffic. We compare this path with the
direct communication path which could have been taken between communication
partners to quantify the impact of \mmss. For this analysis, we 
developed an analysis platform and testbed for mobile applications, called
\matador (Mobile Application Traffic Analysis plattfORm). We use \matador to
exchange messages between a large number of communication partners distributed
over the world using different \mmas and automatically extract
information about the network path the messages used. \section{Related Work}

Several projects worked on analyzing the behavior and communication
patterns of \mmss and the challenges arising when conducting
automated experiments with mobile devices and applications.

Fiadino et al. \cite{fiadino2015vivisecting} investigated characteristics
of WhatsApp communication based on a set of mobile network trace data from
February 2014. In this set, they identified every DNS request to WhatsApp and
resolved them in a distributed way through the RIPE Atlas service. They found
the corresponding address to be exclusively located in the U.S. and
focussed further on Quality of Experience analysis. Huang et al. \cite{huangfine} did
similar work on WeChat, using network traces as well as controlled experiments. For the
latter, they connected phones through WLAN, but relied on heavy manual work for
message sending and traffic analysis. They do not mention a capability to proxy
traffic out through remote nodes. On the collected data, they heavily focus on
dissecting the protocol and architecture. Mueller et al. \cite{mueller2014s}
researched security for a wide set of \mmss and found many
weaknesses, e.g. on the authentication bootstrapping process. They used a 
testbed similar to \matador, but had to explicitly configure the mobile device's proxy
settings. Frosch et al. \cite{froschsecure} provided a detailed protocol
analysis for TextSecure based on its source code.
Anthropological studies like O'Hara et al. \cite{o2014everyday} mainly focus on
the content of messages and their implications on social life, usually featuring
small trace sets or interview data and not data generated from automated testbeds.

The life cycle of network experiments, automated experimentation and testbed
management is in the focus of several related projects. The OpenLab
Project\footnote{\url{http://www.ict-openlab.eu}} focuses on improving network
experimentation for future distributed and federated testbeds and to provide
tools to researchers. Various tools for supporting testbed setup and experimentation
exist~\cite{2015plctools}, but many are outdated or unavailable. None of
these tools support experimentation with mobile devices or geographic diversion of
network traffic.

\cite{2015appTestWiki} provides an extensive list of commercial platforms
aiming to integrate functional mobile application testing in the software
development cycle. Many platforms support the use of real devices and some even
provide testing over mobile carrier networks to ensure functionality. Many
solutions are only provided as a paid service.
 \section{Analyzing Communication of Mobile Messaging Applications}
\label{ref:setup}

In order to analyze the impact of \mmss on traffic locality, our
approach is to compare the \textit{\npath}, defined as direct network path
between communication partners obtained with forward path measurements, and the
\textit{\apath}, defined as the forward path measurements from both
partners to the \mms's backend infrastructure.

We use the \matador testbed to send a large number of messages using different
\mmss between communication partners distributed all over the
globe. To do so, we use \matador equipped with two mobile devices and the
\mma under test. \matador tunnels the application traffic to
PlanetLab nodes as depicted in Figure~\ref{fig:sys}. We
intercept the applications' communication and extract the communication
endpoints. Based on this information, we conduct forward path measurements to
the \mms's backend servers to obtain the \textit{\apath} and
between the nodes to obtain the \textit{\npath}.

We map the hops in both \apath and \npath to countries and
analyze which jurisdictions and political frameworks the traffic traverses on
its way between communication partners. As a result, we can give a qualified
analysis how much the \apath and the \npath differ and if
traffic is confined to a geographic region when both partners are located in
this region.

\subsection{Experimental Setup}

The experimental setup of \matador consists of a dedicated controller node,
two WLAN routers, two Android mobile phones and the PlanetLab proxy nodes as
depicted in Figure \ref{fig:sys}. The controller node orchestrates the overall
experimentation process, configures the WLAN routers, configures the Android
devices and instruments them to send messages. Device instrumentation is
realized using the Android Debug Bridge to configure network connections,
start applications, and issue input events to the devices to automate message
sending. The routers spawn two wireless networks and establish tunnels to the
respective PlanetLab nodes. The router's task is to route, intercept and
modify traffic as well as to automatically parse network traces and start path
measurements to targets. To leverage PlanetLab for this experiment, we use a
tool currently under development at our chair. This tool is able to transparently
proxy traffic over PlanetLab nodes. It is currently in beta status and pending public release.

\vspace{-6mm}
\usetikzlibrary{shapes}
\pgfplotsset{compat=newest}
\usetikzlibrary{shapes,arrows,fit,calc,positioning}
\usetikzlibrary{decorations.pathreplacing}

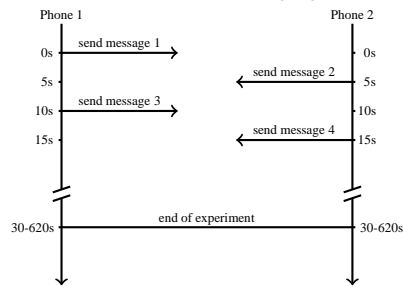
\begin{figure}
\centering

\begin{minipage}{.45\textwidth}
\begin{flushleft}

\caption{Overall experiment design.}
\label{fig:sys}
\begin{scaletikzpicturetowidth}{0.9\linewidth}
\begin{tikzpicture}[scale=\tikzscale]

	\pgfdeclarelayer{bg}    	\pgfsetlayers{bg,main}

\draw [fill=white] (1,2) rectangle (3,4) node [pos=0.5]
	{\parbox{0.5cm}{\centering\tiny Proxy\newline Node}} ;

\draw [fill=white ] (15,2) rectangle (17,4) node [pos=0.5]
	{\parbox{0.5cm}{\centering\tiny Proxy\newline Node}} ;

\draw[fill=white](7.7,7) rectangle (10.3,9) node[pos=0.5]{\tiny Controller};

\draw [fill=white] (4,5) rectangle (6,9) node [pos=0.5] {\tiny Router} ;

\draw [fill=white] (12,5) rectangle (14,9) node [pos=0.5] {\tiny Router} ;

\draw [rounded corners] (6.8,11) rectangle (8,13)   ;
\draw[rounded corners] (6.9,11.4)rectangle(7.9,12.9);
\draw (7.2,11.2)--(7.6,11.2);

\draw [rounded corners] (10,11) rectangle (11.2,13) ;
\draw[rounded corners] (10.1,11.4) rectangle (11.1,12.9);
\draw (10.4,11.2)--(10.8,11.2);

\node  at (1.4,5.6) {\tiny Tunnel} ;
\draw[very thick, dashed] (2,4) to [out=90,in=180] (4,6) ;

\node  at (16.6,5.6) {\tiny Tunnel} ;
\draw[very thick, dashed] (14,6) to [out=0,in=90] (16,4);

\node  at (8.1,9.9) {\tiny USB} ;
\draw[thick] (7.4,11) -- (7.4,9.5) -- (8.7,9.5)--(8.7,9);
\draw[thick] (10.6,11) -- (10.6,9.5) -- (9.3,9.5)--(9.3,9);
\node  at (9.9,9.9) {\tiny USB} ;

\draw[thick](7.7,8) -- (6,8);
\draw[thick](10.3,8)--(12,8);

\draw [fill=black] (6.9,11.1) circle [radius=0.1];

\draw[thick] (6.9,10.8) to [out=180,in=270] (6.6,11.1);
\draw[thick] (6.9,10.6) to [out=180,in=270] (6.4,11.1);
\draw[thick] (6.9,10.4) to [out=180,in=270] (6.2,11.1);

\node  at (5.8,10) {\tiny WLAN} ;
\draw [fill=black] (5,9) circle [radius=0.1];
\draw[thick] (5.3,9) to [out=90,in=0] (5,9.3);
\draw[thick] (5.5,9) to [out=90,in=0] (5,9.5);
\draw[thick] (5.7,9) to [out=90,in=0] (5,9.7);

\draw [fill=black] (11.1,11.1) circle [radius=0.1];
\draw[thick] (11.1,10.8) to [out=0,in=270] (11.4,11.1);
\draw[thick] (11.1,10.6) to [out=0,in=270] (11.6,11.1);
\draw[thick] (11.1,10.4) to [out=0,in=270] (11.8,11.1);

\node  at (12.2,10) {\tiny WLAN} ;
\draw [fill=black] (13,9) circle [radius=0.1];
\draw[thick] (12.7,9) to [out=90,in=180] (13,9.3);
\draw[thick] (12.5,9) to [out=90,in=180] (13,9.5);
\draw[thick] (12.3,9) to [out=90,in=180] (13,9.7);

  \foreach \place/\x in {{(5,4)/1}, {(6.5,2)/2},{(8,2.5)/3},
    {(9.5,1.5)/4}, {(11,4)/5}, {(12.5,2)/6}, {(14,2.8)/7}}
  \node[circle, scale=0.5, fill = black] (a\x) at \place {};

\path[thin] (3,3) edge (a1);
  \path[thin] (a1) edge (a3);
  \path[thin] (a3) edge (a2);
  \path[thin] (a2) edge (a4);
  \path[thin] (a4) edge (a5);
  \path[thin] (a5) edge (a6);
  \path[thin] (a5) edge (a4);
  \path[thin] (a6) edge (a7);
  \path[thin] (a7) edge (15,3);

	\begin{pgfonlayer}{bg}
		\node[cloud, fill=gray!20, cloud puffs=17, cloud puff arc= 100,
        minimum width=5cm, minimum height=1.6cm, aspect=1] at (9,3.5) {\tiny Internet};
\end{pgfonlayer}
\end{tikzpicture}
\end{scaletikzpicturetowidth}
\end{flushleft}
\end{minipage}\hspace{.05\linewidth}\begin{minipage}{.45\textwidth}
	\begin{flushright}
	
	\centering
	\caption{Overview of messaging timers}
	\label{fig:msgtimers}

		\begin{scaletikzpicturetowidth}{0.9\linewidth}
			\begin{tikzpicture}[scale=\tikzscale]
			
			\node (controller) at (0,0.3) {\tiny Phone 1};
			\node (controller) at (10,0.3) {\tiny Phone 2};
			
			\draw[thick] (0,0)--++(0,-5.7);
			\draw[thick](-0.3,-5.8) -- (0.3,-5.6);
			\draw[thick](-0.3,-6.1) -- (0.3,-5.9);
			\draw[->,thick] (0,-6.0)--++(0,-3);
			
			\draw[thick] (10,0)--++(0,-5.7);
			\draw[thick](9.7,-5.8) -- (10.3,-5.6);
			\draw[thick](9.7,-6.1) -- (10.3,-5.9);
			\draw[->,thick] (10,-6.0)--++(0,-3);
			
			\node at (-0.5, -1){\tiny 0s};
			\draw [thick](0,-1) -- (-0.1,-1);
			\node at (10.6, -1){\tiny 0s};
			\draw [thick](10,-1) -- (10.1,-1);
			
			\draw[->,thick](0,-1) -> (4,-1);
			\node at (2,-0.7){\tiny send message 1};
			
			\node at (-0.5, -2){\tiny 5s};
			\node at (10.6, -2){\tiny 5s};
			\draw [thick](0,-2) -- (-0.1,-2);
			\draw [thick](10,-2) -- (10.1,-2);
			
			\draw[->,thick](10,-2) -> (6,-2);
			\node at (8,-1.7){\tiny send message 2};
			
			\node at (-0.6, -3){\tiny 10s};
			\node at (10.5, -3){\tiny 10s};
			\draw [thick](0,-3) -- (-0.1,-3);
			\draw [thick](10,-3) -- (10.1,-3);
			
			\draw[->,thick](0,-3) -> (4,-3);
			\node at (2,-2.7){\tiny send message 3};
			
			\node at (-0.6, -4){\tiny 15s};
			\node at (10.5, -4){\tiny 15s};
			\draw [thick](0,-4) -- (-0.1,-4);
			\draw [thick](10,-4) -- (10.1,-4);
			
			\draw[->,thick](10,-4) -> (6,-4);
			\node at (8,-3.7){\tiny send message 4};

			\draw [thick](0,-7) -- (10,-7);
			\node at (5,-6.7){\tiny end of experiment};
			
			\draw [thick](0,-7) -- (-0.1,-7);
			\draw [thick](10,-7) -- (10.1,-7);
			\node at (-1.0, -7){\tiny 30-620s};
			\node at (11, -7){\tiny 30-620s};

			\end{tikzpicture}
		\end{scaletikzpicturetowidth}

		\end{flushright}
\end{minipage}
\end{figure}
 \vspace{-2mm}

\textbf{Mobile Phone Configuration }To run the \mmas, we use two off-the-shelf, rooted Motorola
Moto-E (2nd generation) smartphones running vanilla Android 5.0.2. For each
device, we created an individual Google Play account. To allow control through
the Android Debug Bridge (ADB), devices are connected to the controller using USB.
We use XPrivacy\footnote{\url{http://repo.xposed.info/module/biz.bokhorst.xprivacy}}
to set the phone's location information according to the location of the specific
PlanetLab node and \texttt{iptables} to restrict network communication to the
specific \mma under test. 
To prevent geolocation based on mobile
network information, the phones were set to
airplane mode with only WLAN enabled.
\textbf{Router Configuration }Two GNU/Linux PCs, configured to act as WLAN access points, provide two
dedicated WPA2-protected wireless networks, one to each mobile phone. Through DHCP,
they provide RFC\,1918 private address and the PlanetLab node's DNS
server to the phones. The routers use \texttt{tcpdump} to intercept traffic and
\texttt{scapy} to automatically process network traces.

\textbf{Measurement Orchestration }The measurements to conduct are defined as \textit{experiments}. Within each
experiment, \matador executes the respective set of commands.  This involves
setting up remote tunnels to two PlanetLab nodes, configuring the network
settings on the routers according to the experiment, starting interception and
manipulation software on the routers, configure the phone to use the wireless
network, setting XPrivacy and firewall settings on the phone, capturing the
phone's screen for later inspection, stepping through the experiment on the
phones with ADB automation, parsing the network trace data automatically, and
executing path measurements to all IP addresses found in the network trace.

\textbf{Experiment Parametrization }To support experimentation with different applications, all required
experimental parameters are controlled through application-specific configuration files.
This includes timers between the different steps of the
experiment, blacklists of hosts not to include in path measurements (e.g. NTP or
DNS servers), the text to send in the messages and how many messages to send
with the application. Such messaging timers, depicted in Figure
\ref{fig:msgtimers}, are controlled through these configuration files.

\textbf{Experiment Monitoring and Error Handling }While running experiments, we learned that using unaltered applications on physical devices 
in this complex setup is prone to errors. We therefore split the overall experiment into smaller junks
to be able to reproduce missing or failing measurements. To be able to detect
and analyze failures, the screen of the mobile devices is captured for each
measurement.

\textbf{Benefits Using the \matador Testbed }
Our approach minimizes effort and cost using common available off-the-shelf
hardware. Since \matador does not rely on device or run time emulation, simulated
network connections, adapted applications, or otherwise modifying the devices in an unusual
way (e.g. setting an application or device proxy), the testbed environment is
transparent to both the phone and apps and looks like a ``normal'' wireless
network. All steps within the experiment life cycle have been automated. This
provides the possibility to efficiently scale the number of
applications and experiments. 
\matador provides functionality to easily and
automatically intercept all network traffic. It can also
transparently redirect network traffic through hosts at remote locations,
appearing to outsiders and the application itself as if the phone was located
at that place.
When proxying the phone's traffic through a remote location,
the phone's location services are manipulated accordingly.
\subsection{Methodology}

The goal of our experiment is to collect information about the path that messages
take on the Internet when two communication partners communicate with each other
using a \mma. In addition, we want to learn about the
regions and countries a message traverses on its way. To do so, we have
to analyze the network path between both communication partners and the messaging
service infrastructure.

In our experiment, we use a set of four carefully selected \mmss and 
use their respective applications to exchange messages between the two mobile
phones in our testbed. In a single measurement, we use one specific \mma, connect
to the \mms on both phones and exchange messages between both
devices. By doing so, we can extract the communication endpoints for the
\mms from the network traffic. We can then perform path
measurements to these communication endpoints from both mobile phones to obtain
the \npath to the service provider infrastructure. To get a global view on
communication, we tunnel traffic through 28 PlanetLab nodes.
This way, we can learn the path messages take for example
for a WhatsApp user in Australia communicating with a user in North America. In
addition, we conduct direct path measurements between both respective PlanetLab nodes to
obtain the direct network path.

For the path measurements, we use the standard \texttt{traceroute} tool provided
with GNU/Linux. From the network traces, we extract the protocol (i.e. TCP or
UDP) and port number (e.g. 443) the \mms uses and apply these
settings to measure the network path to the \mms infrastructure. To
obtain the path between nodes, we use \texttt{traceroute} with TCP and a random
high port.

\textbf{Selection of Applications}
For this work, we carefully selected four different \mmss based on
different characteristics depicted in Table~\ref{tab:app-properties}. 

Based on their popularity, we picked WhatsApp and WeChat as the two \mmss 
built for mobile chat. Due to its high rank in the EFF Scorecard with respect 
to security and privacy and being free software with its source code open to the 
public, we picked TextSecure as a third application for this 
experiment. We chose
Threema for its promise of servers based in Switzerland and claim of
strong privacy for the users. In addition, Threema is one of the few European-based
providers. Since all of the previous solutions rely on a centralized
client/server architecture, we select Bleep as a fifth candidate due to its 
decentralized peer-to-peer architecture. However, we could not enforce peer-to-peer 
behavior in our testbed and observed minute-long delays between messages. We 
concluded that peer-to-peer architectures require closer investigation including  
the use of NAT traversal techniques in our framework. For this reason, we excluded 
Bleep from the set of applications. 

\vspace{-2mm}
	\begin{table}
		\caption{Properties of \mmss and applications.}
	\begin{minipage}{\textwidth}
		\resizebox{\textwidth}{!}
			{\begin{tabular}{lccccc}
    	\toprule
			Application  & Monthly active  & EFF Scorecard\textsuperscript{2} & Architecture & Server  & Primary   \\
			(Version)	  & users\textsuperscript{1}\cite{statista} 	& Points \cite{eff2014}	& 		& Distribution  & mobile\\
			\noalign{\smallskip}
			\hline
			\noalign{\smallskip}

			\cellcolor{black!10}WhatsApp (2.12.176)  & \cellcolor{black!20} 800-900mn \cite{WhatsAppUsersFacebook900,statista2}\cite[p.23]{eufacebookwhatsapp} & 2 & client-server &  n/a & \cellcolor{black!20} \checkmark\\
						\cellcolor{black!10}WeChat (6.2.4)  & \cellcolor{black!20} 400-600mn \cite[p.22]{eufacebookwhatsapp}\cite[p.4]{tencent_report}& n/a  & client-server & n/a & \cellcolor{black!20} \checkmark\\
						Facebook\textsuperscript{3} &\cellcolor{black!10}  350-600mn \cite{facebook_mau},\cite[p.22]{eufacebookwhatsapp} & 2 & client-server & n/a & \xmark\\
						Skype &\cellcolor{black!10}  300mn \cite{skype_mau}  & 1 & client-server & n/a & \xmark\\
						QQ International  &\cellcolor{black!10}  843mn \cite[p.4]{tencent_report} & 2 & client-server & n/a & \xmark\\
						Viber &\cellcolor{black!10}  249mn \cite{vibermau} & 1 & client-server & n/a & ? \\
						LINE  &\cellcolor{black!10}  211mn \cite{line_mau} & n/a & client-server & n/a & ?\\
						Kik  &\cellcolor{black!10}  200mn\textsuperscript{4}\cite{kikmau} & 1 & client-server & n/a & ?\\
						Tango &\cellcolor{black!10}  70mn \cite{tango_mau} & n/a & client-server & n/a & ?\\
						KakaoTalk &\cellcolor{black!10}  48mn \cite{kakaoreport} & n/a & client-server & n/a & ?\\
						Yahoo Messenger& n/a & 1 & client-server & n/a & \xmark\\
			\hline
			\cellcolor{black!10}TextSecure (2.24.1) & \textgreater10mn\textsuperscript{4} \cite{textsecureusers}  & \cellcolor{black!20}7 & client-server & global & \checkmark\\
						Silent Text  & n/a &\cellcolor{black!10}7& client-server & n/a & ? \\
						Telegram  & 30-50mn \cite[p.22]{eufacebookwhatsapp}\cite{telegram_mau}  &\cellcolor{black!10}4\textsuperscript{5}& client-server & global & \checkmark \\
						Wickr   & 4mn\textsuperscript{6} \cite{wickrusers} & \cellcolor{black!10}5& client-server & global & ?\\
			\hline
			\cellcolor{black!10}Bleep (1.0.616)   &  n/a & n/a & \cellcolor{black!20}peer-to-peer & n/a & ?\\
						FireChat &  n/a & n/a & \cellcolor{black!10}peer-to-peer & n/a & \checkmark\\
			\hline
			\cellcolor{black!10}Threema (2.41) & 3mn\textsuperscript{4}\cite{threema_mau} & \cellcolor{black!10} 5 & client-server & \cellcolor{black!20} Switzerland & \checkmark\\
						SIMSme & 1 mn\textsuperscript{6}  & n/a & client-server & \cellcolor{black!10}  Germany & \checkmark \\
			\hline
			\multicolumn{6}{l}{1: Around July 30, 2015, for exact date see app-specific source  2: EFF Secure Messaging Scorecard \cite{eff2014}} \\
			\multicolumn{6}{l}{3: Stand-alone Facebook Messenger  4: Registered users 5: Score of 7 in secure chats  6: App Store Downloads} \\
						\bottomrule
		\end{tabular}}
		\end{minipage}
		\label{tab:app-properties}

	\end{table}
 \vspace{-2mm}

\textbf{Node Selection }To achieve a global view on messaging communication, we compiled a list of
PlanetLab nodes providing a wide geographical distribution. The objective for this list was
to cover as many regions and countries as possible. However, PlanetLab does not
provide equal coverage in all regions and availability of nodes strongly differs
across regions. When we conducted our experiment, PlanetLab featured nodes
in 49 countries, but we only found 28 countries with at least one stable and
responsive node, providing good coverage for North America,
Europe, Asia and Oceania. For South America only a single node in Argentina and 
Brazil was provided, for Africa no nodes could be accessed at all.

For our experiment, we therefore used 4 nodes in the Americas (North America: 2,
South America: 2), 7 nodes in Asia (Eastern Asia: 4, South-Eastern Asia: 2,
Western Asia: 1), 16 nodes in Europe (Eastern Europe: 3, Northern Europe: 5,
Southern Europe: 4, Western Europe: 4) and 2 nodes in Oceania.

\textbf{Limitations }It is important to note that our path
measurements only record a country as being part of a path if a hop from that
country replies to path measurements. This can be biased by (a)
nodes not answering those requests and (b) countries being passively traversed.
Especially the latter is relevant, as intelligence services are known to also
wiretap passively. For example, some measurements from Switzerland offer
direct paths to Hong Kong or the U.S., but obviously more countries in between
would have passive access to the cables in-between. \section{Postprocessing Experiment Results}Despite limiting application communication, the resulting network traces included 
some irrelevant flows.
For this experiment, we solely want to evaluate
traffic between the \mma and the \mms's backend.
Therefore, we had to classify network flows and assemble a black- and
whitelist of network flows for exclusion or inclusion. Here, we went through several steps:

First, we limited background traffic by firewalling communication to only allow
\mma to access the network.Second, we conducted six measurements from America, Europe and Asia
without the \mma running. This resulted in network traces containing
``background noise'' we could exclude after manual validation.Third, we manually inspected several dozens of traces per \mma to determine additional
background traffic. The sources for this traffic were manually added to the
filtering blacklist.Fourth, we separated authentication and other background
traffic for every application from messaging traffic through temporal
correlation with message timers.

For Threema, TextSecure and WhatsApp, we found all messaging servers to be
resolved through DNS and to resolve uniformly across the globe, confirming the
results of \cite{fiadino2015vivisecting} for WhatsApp. We found WeChat to use
both DNS requests and a custom-built DNS-over-HTTP protocol for name resolution,
providing different name resolution when queried from within or from outside
China. This DNS-over-HTTP uses a 30-minute timeout and therefore ``contaminates''
our name resolution cache, which we flush after every experiment, typically lasting five to ten minutes.
We therefore built
the whitelist for WeChat analysis through manual analysis. The resulting
detailed DNS table can be found online\footnote{\url{http://www.net.in.tum.de/pub/mobmes/dnstable.pdf}}.

In a last step, we automatically processed all traces and classified all
addresses into this black- or whitelist. For remaining addresses, we manually
classified them along the observed traces and re-ran the classifier until all
addresses could be classified.

\subsection{Mapping Path Measurements to Countries and Regions}

To obtain the countries the traffic traverses, both the \apath and the
\npath were processed to provide a geolocation of the IP addresses.
With some manual corrections, we found the
ip2location\footnote{\url{http://www.ip2location.com}} country
database to provide the most accurate results. To not overly rely on that
database, we manually validated the mappings in at least one trace per target
subnet and source country.  With respect to known inaccuracies of both reverse
DNS labels and geolocation databases, as described in
\cite{huffaker2014drop,zhang2006dns}, we paid special attention to round-trip
times found in forward path measurements.

To analyze locality with respect to a specific geographic region, we used the
United Nations
geoscheme\footnote{\url{http://millenniumindicators.un.org/unsd/methods/m49/m49regin.htm}}
to assign countries to regions and subregions. This scheme relies on 5 regions
(Africa, Americas, Asia, Europe, Oceania) which are further divided into geographic
subregions (e.g. for the Americas: Latin America and the Caribbean, Central America, 
South America, and Northern America).\subsection{Mapping Countries to Interest Groups}

In addition to geographic locality, we analyzed the possibility of
several jurisdictions and similar entities to access the network traffic. In
this analysis, we defined several \textit{interest groups} and checked for the
different \mmss if these interest groups can access the traffic.
For this analysis we defined the following interest groups:

\vspace{-1mm}
\begin{itemize}
  \item {\textit{5 Eyes}} consisting of: Great Britain, United States,
    New Zealand, Canada
  \item {\textit{European Union}} consisting of: Austria, Belgium, Bulgaria,
  Croatia, Cyprus, Czech Republic, Denmark, Estonia, Finland, France, Germany, Greece,
  Hungary, Ireland, Italy, Latvia, Lithuania, Luxembourg, Malta, Netherlands,
  Poland, Portugal, Romania, Slovakia, Slovenia, Spain, Sweden, United Kingdom
  \item {\textit{Arab League}} consisting of: Algeria, Bahrain, Comoros,
  Djibouti, Egypt, Iraq, Jordan, Kuwait, Lebanon, Libya, Mauritania, Morocco,
  Oman, Palestine, Qatar, Saudi Arabia, Somalia, Sudan
  \item {\textit{Russia}} with the only member Russia
  \item {\textit{China}} with the only member China
\end{itemize}
 \section{Results}

With our experiments, running from Sep 30 2015 to Oct 12 2015, we conducted
406 measurements between the 28 PlanetLab nodes using the 4 selected \mmss,
resulting in 1624 measurements in total.

\begin{table}
  \caption{Geographical traffic locality in regions for {\npath}s and {\apath}s.}
  \label{tab:results_regions}
	\begin{minipage}{\textwidth}
		\resizebox{\textwidth}{!}
    {\begin{tabular}{lrrrrrrrr}
    \toprule
        & & \multicolumn{6}{c}{Traffic leaving region} \\
        \cmidrule{3-9}
    &   & \multicolumn{2}{c}{\NPath}& & \multicolumn{4}{c}{\APath}\\
    \cmidrule{3-4} \cmidrule{6-9}
    Region & \# Measurements  & \ \ \ \ \ \ \ \ \  \#  & \%  & &TextSecure & Threema & WeChat & WhatsApp \\

    \midrule

Europe & 120 & 0 & 0\% & &   100\% &  \textbf{0\%} & 100\% & 100\% \\

Oceania & 3 & 0 & 0\% &  &  100\% & 100\% & 100\% &  100\% \\

Asia & 28 & 6 & \textbf{21\%} &&  100\% &  100\% & \textbf{50\%} &  100\% \\
Americas & 10 & 0 & 0\% &&  0\% & 100\% & 100\% &  0\% \\

\midrule
\ \ South America & 3 & 1 & \textbf{33\%} & & 100\% & 100\% & 100\% & 100\% \\
\ \ Northern America & 3 & 0 &0\%  & & \textbf{0\%} & 100\% & 100\% & \textbf{0\%} \\
\bottomrule 

    \end{tabular}}
  \end{minipage}
\end{table}
 
Table~\ref{tab:results_regions} shows the path comparisons between \apath
and \npath. The first columns evaluate the direct measurements between nodes
and show how many \% of measurements failed to stay within the region. We found that
all traffic from Israel to other Asian countries is being routed
through Europe and the U.S. As we use seven nodes in Asia, six measurements from
Israel fail to remain within region. Also, with two nodes in South America, the
measurement between those two nodes leaves South America for routing through North
America. As the two in-country measurements stay in the region, the
33\% understate the effect, caused by the low number of nodes.  As a result we highlight
that only Europe and North America feature at least one messenger that
keeps traffic local. Asia traffic for WeChat does not remain local because of
Israel's aforementioned routing and also because of traffic from Singapore and
Thailand being routed to the Chinese WeChat servers through U.S. IXPs.

Table \ref{tab:geolocapps_jur} shows how measurements from a
specific region were subject to various interest groups, both for the
\npath and for the specific \apath:

\textbf{Europe to Europe:} 72\% of
\npath measurements within Europe were accessible to 5 Eyes (by routing
through UK). 98\% of measurements were accessible to the European Union, with
only measurements internal to Switzerland and Norway not being accessible. For {\apath}s,
Threema reduces the 5 Eyes access by 16\% as it effectively proxies traffic
through Switzerland, which enforces continental routing for some routes (e.g.
Poland - Switzerland - Spain as compared to Poland - UK - Spain). 99\% of WeChat
measurements within Europe were accessible to 5 Eyes because of routing
through the U.S. Only
the Switzerland internal measurement offered a direct path to Hong Kong.
As Switzerland has a direct path to the U.S. as well, this also explains the one
case where EU can not access TextSecure messages. When using Threema within
Switzerland, the \apath remains in Switzerland as well, hence the EU
cannot access those measurements.	

\textbf{Oceania to Oceania:} As Australia and New Zealand are both members of 5
Eyes, obviously all measurements are accessible to the latter. It is remarkable
that all WeChat traffic, e.g. generated by exile Chinese, is routed through
China.

\textbf{Asia to Asia:} At a network level, both 5 Eyes, China and the
European Union can access about 20\% to 40\% of traces sent within Asia. This is
largely caused by the before mentioned Israel routing. 75\% of Threema traffic
is 5 Eyes accessible by routing to Switzerland through the U.S. Also, a large
portion of WeChat traffic (46\%) is accessible to 5 Eyes, both by Israel routing
through the U.S. and by Singapore routing to WeChat's Chinese backend through an 
U.S. IXP.

\textbf{North America to North America:} As expected, 100\% of traffic is 5 Eyes
accessible. For Threema, traffic from Canada to Switzerland was again routed through
a direct hop from Miami to Zurich, resulting in two measurements seeming
inaccessible to EU.

\textbf{South America to South America:} Measurements from Argentina were routed
through a direct tunnel from Miami to Zurich and hence were not accessible for the EU in our
metric. Hence only 2 out of 3 Threema measurements from South America are 
accessible for the EU. However, South America's communication is, independently of the
\mms being used, always susceptible to 5 Eyes.

\textbf{Russia and Arab League:} None of the measurements did traverse Russia or
the Arab League. We hence excluded those from the table.

\begin{table}
 \setlength{\tabcolsep}{5pt}
\caption{Accessibility of traffic for different jurisdictions.}
\label{tab:geolocapps_jur}
\begin{minipage}{\textwidth}
  \resizebox{\textwidth}{!}
  {\begin{tabular}{llrrrrrrrrrrrr}
    \toprule
    & & & \multicolumn{9}{c}{Accessible for Jurisdiction} \\
    \cmidrule{4-13}
    & Juris- &  & \multicolumn{2}{c}{\NPath} &\multicolumn{2}{c}{TextSecure}&\multicolumn{2}{c}{Threema}&\multicolumn{2}{c}{WeChat}&\multicolumn{2}{c}{WhatsApp}\\
    \cmidrule{4-5}\cmidrule(lr){6-7}\cmidrule{8-9}\cmidrule(lr){10-11}\cmidrule{12-13}
    Region & diction  & \#Total  & \# & \%  &  \# & \% & \# & \% & \# & \% & \# & \% \\
    \midrule
Europe & 5 Eyes & 120 & 86 & \textbf{72\%} &  120 & 100\% & 68 & \textbf{57\%} & 119 & 99\%  & 120 & 100\%  \\
Europe & EU & 120 & 118 & 98\% &  119 & 99\% & 119 & 99\% & 120 & 100\%  & 120 & 100\%  \\
Europe & China & 120 & 0 & 0\% &  0 & 0\% & 0 & 0\% & 120 & 100\%  & 0 & 0\%  \\
\midrule
Oceania & 5 Eyes & 3 & 3 & 100\% &  3 & 100\% & 3 & 100\% & 3 & 100\%  & 3 & 100\%  \\
Oceania & EU & 3 & 0 & 0\% & 0 & 0\% & 3 &  100\% & 0 & 0\%  & 0 & 0\%  \\
Oceania & China & 3 & 0 & 0\% &  0 & 0\% & 0 & 0\% & 3 & 100\%  & 0 & 0\%  \\
\midrule
Asia & 5 Eyes & 28 & 6 & \textbf{21\%} &  28 & 100\% & 21 & \textbf{75\%} & 14 & \textbf{50\%}  & 28 & 100\%  \\
Asia & EU & 28 & 6 & \textbf{21\%} &  7 & 25\% & 18 & 64\% & 7 & 25\%  & 7 & 25\%  \\
Asia & China & 28 & 10 & 36\% & 7 & 25\% & 7 & 25\% & 28 & 100\%  & 7 & 25\%  \\

\midrule
South America & 5 Eyes & 12 & 4 & \textbf{33\%} & 3 &  100\% & 3 & 100\% & 3 & 100\%  & 3 & 100\%  \\
South America & EU & 12 & 0 & 0\% &  0 & 0\% & 2 & 67\% & 0 & 0\%  & 0 & 0\%  \\
South America & China & 12 & 0 & 0\% & 0 & 0\% & 0 & 0\% & 3 & 100\%  & 0 & 0\%  \\
\midrule
North America & 5 Eyes & 12 & 12 & 100\% & 3 & 100\% & 3 & 100\% & 3 & 100\%  & 3 & 100\%  \\
North America & EU & 12 & 0 & 0\% & 0 & 0\% & 2 & 67\% & 0 & 0\%  & 0 & 0\%  \\
North America & China & 12 & 0 & 0\% & 0 & 0\% & 0 & 0\% & 3 & 100\%  & 0 & 0\%  \\
\bottomrule

  \end{tabular}}
\end{minipage}
\end{table}
  \section{Summary and Conclusion}

We conducted traffic locality measurements between 28 countries for four
\mmss. We found those apps to heavily distort locality of traffic and
hence drastically widen the set of actors able to access it. With a few
notable exceptions, e.g. when using Threema in Switzerland, this has large
negative impacts on the users' privacy. With this being the first study on this
particular topic, we hope to raise user and operator awareness to the problem at
hand. To conduct our measurements, we introduced the \matador framework to analyze 
messaging traffic characteristics on mobile phones. A detailed overview over the
\matador framework can be found in~\cite{zirngibl2015}. We fully release both
the \matador framework and the dataset produced in our measurements through our
website\footnote{\url{http://net.in.tum.de/pub/mobmes/}}. This
enables future work to easily validate our results or do further analysis, such
as deeper protocol analysis on the apps. Future work might also include analysis
of WeChat's regional optimization within China, focus on restricted end user connectivity to be able to conduct a similar study with
peer-to-peer services like Bleep, or further dissecting protocols of
other \mmss.
\textbf{Acknowledgments: } We thank Andreas Loibl for early access to
his Measurement Proxy software.
 
\apptocmd{\sloppy}{\hbadness 10000\relax}{}{}
\apptocmd{\thebibliography}{\raggedright}{}{}
\bibliography{PAM16_31-scheitle}
\bibliographystyle{abbrv}

\end{document}